\documentclass[aip,cha,amsmath,amssymb,reprint,a4paper]{revtex4-1}
\usepackage{graphicx}
\usepackage{dcolumn}
\usepackage{bm}
\usepackage[colorlinks=true,allcolors=blue]{hyperref}
\usepackage{times}
\usepackage{siunitx}
\usepackage{chemformula}

\begin{document}

\title{Multiplexed random-access optical memory in warm cesium vapor}



\author{Leon Meßner}
\email{messner@physik.tu-berlin.de}
\affiliation{Institut f\"ur Physik, Humboldt-Universit\"at zu Berlin, Newtonstr. 15, 12489 Berlin, Germany.}
\affiliation{Deutsches Zentrum f\"ur Luft- und Raumfahrt e.V. (DLR), Institute of Optical Sensor Systems,  Rutherfordstr. 2, 12489 Berlin, Germany.}

\author{Elizabeth Robertson}
\affiliation{Deutsches Zentrum f\"ur Luft- und Raumfahrt e.V. (DLR), Institute of Optical Sensor Systems,  Rutherfordstr. 2, 12489 Berlin, Germany.}
\affiliation{Technische Universität Berlin, Institut für Optik und Atomare Physik, Str. des 17 Juni 135, 10623 Berlin, Germany}
\author{Luisa Esguerra}
\affiliation{Deutsches Zentrum f\"ur Luft- und Raumfahrt e.V. (DLR), Institute of Optical Sensor Systems,  Rutherfordstr. 2, 12489 Berlin, Germany.}
\affiliation{Technische Universität Berlin, Institut für Optik und Atomare Physik, Str. des 17 Juni 135, 10623 Berlin, Germany}
\author{Kathy Lüdge}
\affiliation{Technische Universität Ilmenau, Institut für Physik, Weimarer Straße 25, 98693 Ilmenau, Germany}
\author{Janik Wolters}
\affiliation{Deutsches Zentrum f\"ur Luft- und Raumfahrt e.V. (DLR), Institute of Optical Sensor Systems,  Rutherfordstr. 2, 12489 Berlin, Germany.}
\affiliation{Technische Universität Berlin, Institut für Optik und Atomare Physik, Str. des 17 Juni 135, 10623 Berlin, Germany}

\date{\today}
  
\begin{abstract}
  The ability to store large amounts of photonic quantum states is regarded as substantial for future optical quantum computation and communication technologies.
  However, research for multiplexed quantum memories has been focused on systems that show good performance only after an elaborate preparation of the storage media.
  This makes it generally more difficult to apply outside a laboratory environment.
  In this work, we demonstrate a multiplexed random-access memory to store up to four optical pulses using electromagnetically induced transparency in warm cesium vapor.
  Using a $\Lambda$-System on the hyperfine transitions of the Cs D1 line, we achieve a mean internal storage efficiency of 36\% and a $1/\textrm{e}$ lifetime of 3.2~µs.
  In combination with future improvements, this work facilitates the implementation of multiplexed memories in future quantum communication and computation infrastructures.

\end{abstract}
\maketitle
\onecolumngrid

\section{Introduction}
Quantum memories are considered to be a main component for the realization of many future second generation quantum technologies.
Their potential use ranges from synchronizing inputs into various types of quantum systems~\cite{Heshami.2016} to re-configurable optical reservoir computing~\cite{Jaurigue.2021b}.
They enable on-demand operation of otherwise probabilistic single-photon sources and quantum gates~\cite{Poem.2018}, which will significantly enhance their rate of operation~\cite{Bhaskar.2020}.
Moreover, they have been identified as an essential device required to realize a quantum repeater~\cite{Briegel.1998}, a key technology needed for long-distance quantum communication.
When specifically considering the implementation of a global quantum communication network, satellite based quantum communication has been hallmarked as a most promising system if enhanced with a multiplexed quantum memory.
It has been shown that a significant increase in communication rate is already achievable with around 1000 randomly accessible storage modes~\cite{Gundogan.2021,Wallnofer.2021}.
Consequently, the realization of suitable multiplexed quantum memories will be an important milestone in extending quantum communication over long distances.

Quantum memories have been demonstrated using a variety of storage protocols in a number of single emitter and ensemble-based matter systems.
These include solid state systems, ultra-cold atoms and warm atomic vapors~\cite{Heshami.2016}.
Although the routine formation of subnanokelvin Bose-Einstein condensates in earth's orbit has been demonstrated~\cite{Aveline.2020} and is pursued in future projects~\cite{Frye.2021}, reducing the technological requirements for space-borne quantum memories is a key step.
This makes memories based on warm atomic vapors favored for applications, as they require no vacuum, laser cooling or strong magnetic fields.

The memory used in this experiment utilizes the effect of electromagnetically induced transparency (EIT) on the $\Lambda$-system composed by the $6^2S_{1/2} F{=}3, F{=}4$ and the $6^2P_{1/2} F{=}3$ atomic hyperfine levels of an ensemble of cesium atoms to map optical excitations to a long-lived spin-wave, i.e. a coherence of the two hyperfine ground states of the atomic ensemble~\cite{Imamoglu.2005,Gorshkov.2007b}.
Due to its coherent ensemble origin, this spin-wave shows comparatively low dephasing and loss.
Subsequent retrieval of the spin-wave excitation into the input optical mode can then be performed at some chosen time that is smaller than the spin-wave lifetime.
Light storage for up to 1~s~\cite{Katz.2018} and single-photon operation~\cite{Esguerra.2022,Buser.2022} have been demonstrated in separate experiments in warm atomic vapors.

For a quantum memory to be most useful within a quantum communication system, the memory must be scaleable with the possibility to access individual storage modes in a way that is not significantly limited by the used technology.
Various forms of quantum memory multiplexing have been studied in the past, including time bin~\cite{Usmani.2010}, orbital angular momentum~\cite{Ding.2015}, and spatial~\cite{Langenfeld.2020,Li.2021} multiplexing.
Among these approaches, Ref.~\cite{Li.2021} shows a clear foreseeable path to achieving the required number of 1000 randomly accessible modes.
However, the technological overhead of a cold atom setup complicates operating these outside of a laboratory.
In this work we demonstrate a memory that combines the advantages of using warm vapor with a path towards scaleable multi-rail operation.

\section{Experiment}
\begin{figure}
\centering\includegraphics[width=\textwidth]{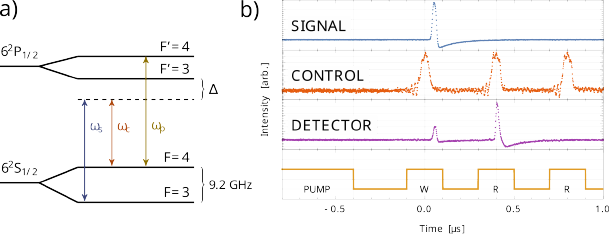}
\caption{Memory scheme. a) Scheme of the Cs D1 energy levels' hyperfine structure that forms the $\Lambda$-system used. Signal and control fields are red detuned from resonance by $\Delta$ and have angular frequencies $\omega_\textrm{s}$ and $\omega_\textrm{c}$ respectively. The transition $F=4 \rightarrow F'=4$ is resonantly driven by the pump field with angular frequency $\omega_\textrm{p}$. b) Time traces from an experiment for writing a signal pulse onto an atomic spin-wave at time $t_0=0$ and retrieving it at $t_0 + 0.4~\textrm{µs}$. The bottom panel shows the operation performed on the memory (W: write, R: read).}\label{fig:lambda_and_pulsetrain}
\end{figure}

The multi-rail memory presented here uses an EIT memory scheme~\cite{Fleischhauer.2005} on the hyperfine ground state transitions of the Cs D1 line, as shown in Fig.~\ref{fig:lambda_and_pulsetrain}(a).
The control and signal lasers (14-pin butterfly external cavity diode laser (ECDL) modules by Sacher Lasertechnik) are set on the $F{=}4 \rightarrow F'{=}3$ and the $F{=}3 \rightarrow F' {=} 3$ transitions respectively.
To stabilize the frequency difference between signal and control laser to the Cs ground state hyperfine splitting, light from both lasers is superimposed on a fast photodiode (Electro Optics Technology, ET-3500FEXR) and their beat frequency is offset locked~\cite{Schunemann.1999} using a RedPitaya FPGA-board running the Linien~\cite{Wiegand.2022} locking software.
 We generate Gaussian signal and control pulses with a full width at half maximum (FWHM) of 25~ns and 43.75~ns respectively using an arbitrary function generator (AFG, Tek AFG31152).
These pulses are modulated onto cw laser beams with electro-optic amplitude modulators (EOMs, Jenoptik AM905).
The experiment is designed such that the signal and control pulses have linear and orthogonal polarization to each other to reduce the control light leaking into the detection path.

The atomic storage medium is confined to a cylindrical, 25x75~mm anti-reflection coated cesium vapor cell filled with 5 torr \ch{N2} of buffer gas.
It is kept at 60°C and shielded from ambient magnetic fields by a double-layered mu-metal housing.
Spatial addressing of different rails within the cell is performed by one acousto-optic deflector (AOD) in front and one behind the vapor cell.
The AODs (AA MT200-B100A0,5-800) have an aperture of 0.5x2 $\textrm{mm}^2$ and a measured deflection of 0.2~mrad/MHz.
Each of them is driven by the frequency sum of an arbitrary function generator (AFG, Tek AFG31152) and a local oscillator (Mini-Circuits, ZOS-300+) resulting in 200$\pm$50~MHz of carrier frequency.
We refer to the position of memory rails by the AOD driving frequency used to deflect the beam to that position; the distance between rails is thus expressed as the difference in driving frequency.
Changing the AOD driving frequency by 8~MHz changes the lateral position of the deflected beam by 270~µm, equaling one signal beam radius.
\begin{figure}
\centering\includegraphics[width=\textwidth]{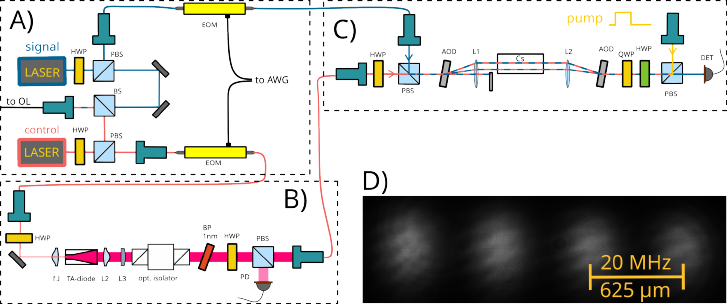}
\caption{Sketch of the experiment with A) laser sources, spectroscopy and optical pulse shaping, B) TA based pulse amplification, C) multi-rail storage system and D) CCD image of the four used rails, with the camera at the place of the Cs cell. HWP: half-wave plate, QWP: quarter-wave plate, DET: detector, (P)BS: (polarizing) beam splitter, L1;L2/L3: aspheric/cylindrical lens, AOD: acousto-optic deflector, EOM: electro-optic modulator, AFG: arbitrary function generator, IF: interference filter, OL: offset lock. }\label{fig:setup}
\end{figure}

While the signal pulses enter the memory unmodified after their generation, the control laser pulses are amplified by a self-made tapered amplifier (TA, see~\cite{Schkolnik.2017}) and then spectrally filtered by a dielectric bandpass interference filter (IF) with a 1~nm FWHM.
This results in 200~mW of coupled cw power.
To increase the control laser's on-off ratio, the TA diode's driving current is only switched on for 120~ns, centered on the optical pulse, using a 4~A dc-coupled input follower driver with 2~ns rise/fall time.
This reduces the unwanted interaction between control laser and atoms in times when no control pulse is generated.

The beam paths of the cross-polarized signal and control lasers are overlapped at a polarizing beam splitter (PBS) on one side of the memory and then propagate collinearly through the cesium cell, and both AODs.
At the position of the cell, control and signal beam have a $1/e^2\textrm{-level}$ radius of 350~µm and 270~µm respectively.
This yields an atomic transit time of $\Delta t=3.7~\textrm{µs}$ for one signal beam radius when using a 2D diffusion model of $\Delta x = \sqrt{4 D \Delta t}$ for the diffusion length and an assumed diffusion constant of $D_0=0.24~\textrm{cm}^2\textrm{s}^{-1}$~\cite{Thomas.2017} at $T_0=0~\textrm{K}$ and $P_0=760~\textrm{torr}$.
After traversing the second AOD, the signal and control beams are split by a second PBS and are then individually coupled to fibers and detected by either a Si photodiode (Thorlabs DET10A2) or a Si avalanche photodiode (Menlo Systems APD210).

Optical pumping of the \ch{Cs} atoms into the $6^2S_{1/2}F{=}3$ state is performed by a third ECDL laser locked to the $6^2S_{1/2}F{=}4\rightarrow 6^2P_{1/2}F{=}4$ transition by saturated absorption spectroscopy.
The pump light power is controlled via transmission through an electrically pulsed semiconductor optical amplifier (SOA) and illuminates each memory rail with 20~mW of optical power for 900~ns prior to the memory experiment sequence.

Figure~\ref{fig:lambda_and_pulsetrain}(b) shows a typical time trace for a single-rail storage experiment and a sketch of the experimental setup can be seen in Fig.~\ref{fig:setup}.
Several features of storage within an EIT medium can be observed in the time trace.
At $t=0~\textrm{µs}$ a signal pulse enters the atomic medium and is partly mapped to an atomic spin wave by the control laser field.
The portion of that signal pulse that is transmitted through the atomic vapor is detected by the photodiode as leakage.
After $0.4~\textrm{µs}$ the control laser field is switched on again and retrieves the spin-wave excitation back into the signal beams optical mode.
A third pulse of the control laser at $t=0.8~\textrm{µs}$ serves to determine if all the excitation has been retrieved and also allows to estimate the signal noise induced by the control laser field.
Not having a significant detection event during this last pulse, we conclude that nearly all the spin-wave excitation is mapped backed to optical and signal to noise ratio is not a limiting factor for this experiment with aforementioned laser pulses.

\section{Results}
Prior to performing multi-rail storage, we first identify optimal operating conditions for the multi-rail memory by assessing the influence of rail separation on the interactions between two memory rails, and subsequently minimizing the cross-talk.
For comparison of the single and multi-rail operation, we measure the 1/e lifetime and memory efficiency per rail.
  
To assess the influence of rail separation on their interaction, multiple storage experiments at different rail separations are conducted.
For effective operation of a memory, we require that operations on a given memory rail do not affect its neighbors.
To determine the minimal separation that shows no cross-talk, we write into a rail fixed at 190~MHz, read from a neighboring rail, and then read from the 190~MHz rail again.
\begin{figure}
  \centering\includegraphics[width=0.8\textwidth]{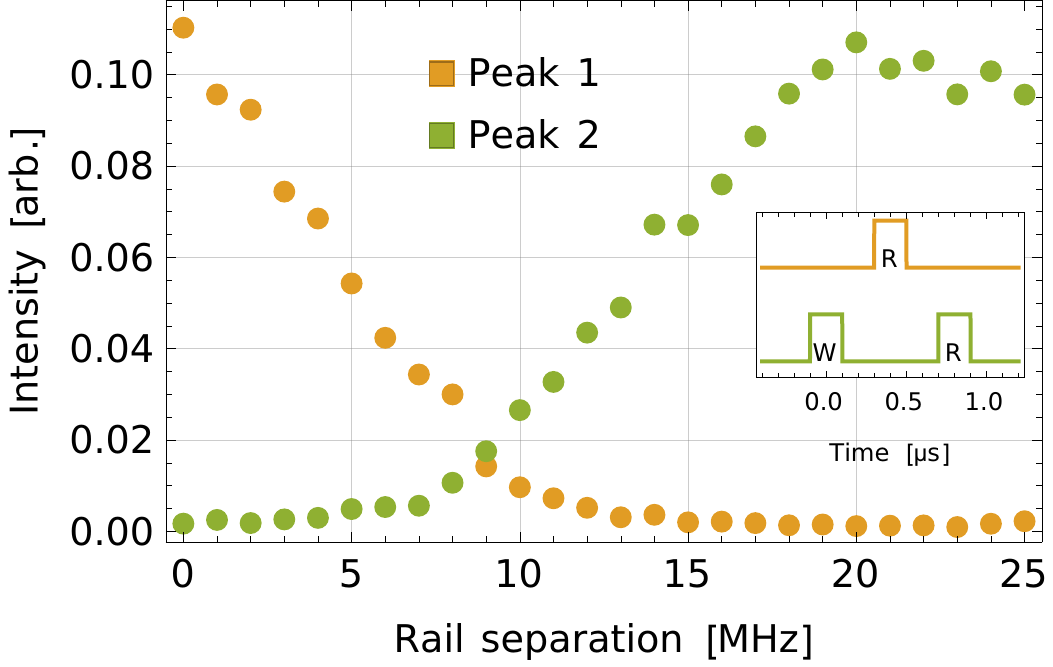}
\caption{Cross-talk estimation. Intensities detected in the first (Peak 1, orange rail) and second (Peak 2, green rail) read peak depending on the rail separation for the experiment sequence depicted in the inset. For a difference of 20~MHz in AOD driving frequency (rail separation), the influence between neighboring rails vanishes. The inset shows the used experiment sequence consisting of a write on the green rail at $t=0$, a read on a neighboring orange rail at $t=0.4~\textrm{µs}$ and finally a read on the green rail at $t=0.8~\textrm{µs}$.}\label{fig:crossread}
\end{figure}
This  \textbf{w}rite/\textbf{r}ead/\textbf{r}ead sequence is depicted in the inset of Fig.~\ref{fig:crossread}.
The rail separation is varied from 0 to 25~MHz at steps of 1~MHz, and the retrieval peak intensities after the first and second read are measured.
The results are depicted in Fig.~\ref{fig:crossread}.
Below 5 to 8~MHz of separation no excitation is left for the second retrieval pulse and both read pulses address the same ensemble of atoms.
At a separation of 20~MHz the influence of the read operation on the neighboring rail is no longer visible.

The AOD device used has a 100~MHz bandwidth and a 25\% reduced diffraction efficiency at the edges of the frequency range; consequently we chose to limit this experiment to four memory rails spaced by 20~MHz.
A CCD image of the four rails, taken at the position of the Cs cell, is shown in Fig.~\ref{fig:setup}(D).
A straightforward method to increase the number of rails, is to use AODs with a higher number of resolvable spots.

\begin{figure}
\centering\includegraphics[width=0.8\textwidth]{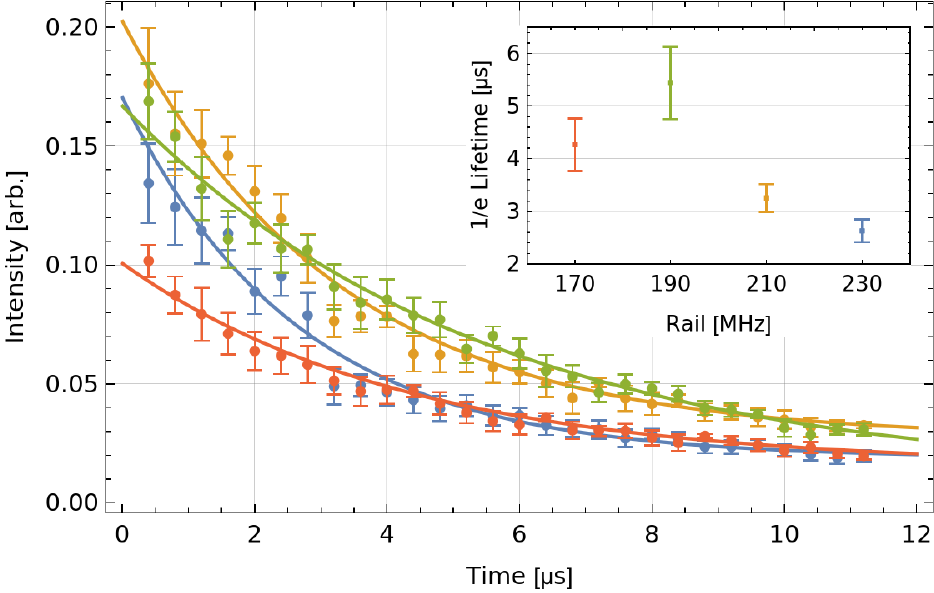}
\caption{Measured retrieval amplitudes for storage times between 0.4 and 11.4~µs together with an exponential fit to the values. The inset shows the per rail 1/e storage lifetime deduced from the fit.}\label{fig:lifetimeplot}
\end{figure}

The 1/e storage lifetime per rail is determined by performing storage experiments with increasing time delay between the memory write and read operation, for each rail.
The delay was varied between 0.4~µs and 11.2~µs in steps of 400~ns, and for each delay, we measure the retrieved peak intensities, averaged over 500 repetitions.
Uncertainties are given by the standard deviation of the intensities.
The intensities were fitted with an exponential function to extract the 1/e lifetime.
Measured retrieval peak intensities and fit function are displayed in Fig.~\ref{fig:lifetimeplot}.

\begin{table}[b]
  \centering
  \begin{tabular}{|l || r | r | r | r | r |}
    \hline
    Rail (MHz) & 170 & 190 & 210 & 230 & Mean\\
    \hline
    Lifetime (µs) & 4.3(5) & 5.4(7) & 3.3(3) & 2.6(3) & 3.2(2)\\
    Efficiency (\%) & 32 & 35 & 39 & 36 & 36\\
    \hline
  \end{tabular}
  \caption{Measured 1/e-lifetime and retrieval efficiency for each rail and weighted mean.}
  \label{tab:lifetime_efficiency}
\end{table}

The resulting lifetime values per rail are shown in Table~\ref{tab:lifetime_efficiency} together with the achieved internal memory efficiencies $\eta_{\textrm{mem}}$ at $t=0$.

Since the measured lifetimes are consistent with the estimate using the simple diffusion model presented earlier, it is reasonable to assume that diffusion is the most important lifetime-limiting process for the beam diameters chosen in this work.

Independent investigation on a single rail setup also showed that spin polarization lifetimes at least on the order of several hundred microseconds are possible with larger beam diameters.

The memory efficiencies are calculated by extrapolating the pulse energy of a retrieved pulse after $t=0~\textrm{µs}$ of storage from a retrieved pulse after $t=0.4~\textrm{µs}$ of storage using the memory lifetime.
This is then divided by the energy of a normalization pulse to yield the efficiency.
To obtain the normalization pulse, we set the signal laser frequency 2~GHz below the $F{=}3\rightarrow F'{=}3 $ transition frequency, block the control beam and record the transmitted signal pulse.
Under these conditions, we assume the pulses not to be absorbed by the atoms.

\begin{figure}
  \centering
  \includegraphics[width=0.8\textwidth]{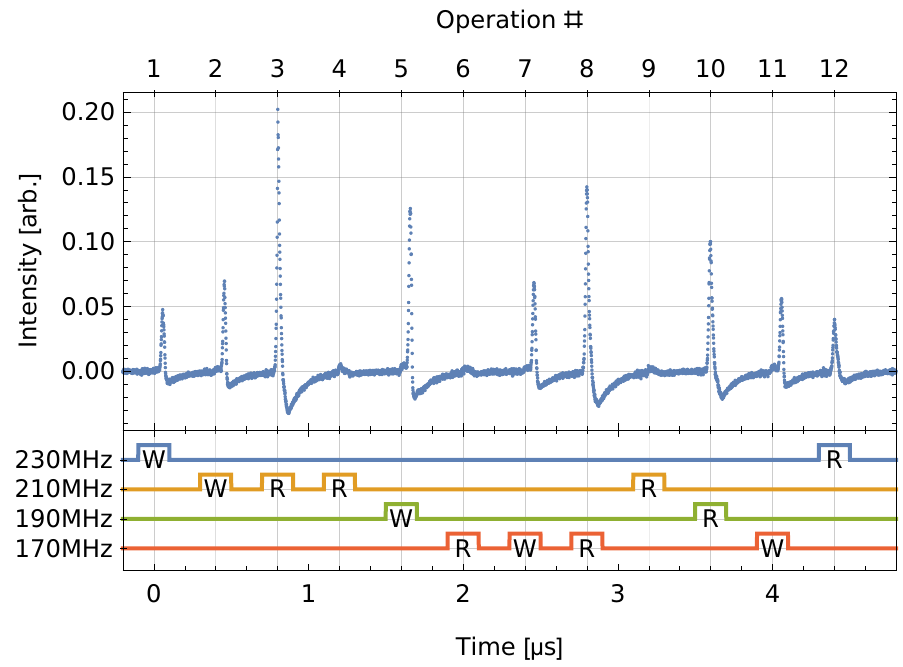}
  \caption{Storage experiment in the random-access memory using four rails with detected intensity in the signal path in the top and performed operations (\textbf{r}ead/\textbf{w}rite) in the bottom panel. A total of 12 operations are performed over a span of 5~µs.}
  \label{fig:rnd-plot}
\end{figure}

Using the insights and results from the measurements on lifetime, efficiency and rail separation, we now explore the possibility of random-access operation in the memory setup.
For this purpose an experimental sequence was designed that highlights important criteria for use as a random-access quantum memory.
Figure~\ref{fig:rnd-plot} illustrates this experiment.

The bottom panel depicts the operation performed on each specific rail and the top panel shows the intensity detected by the APD over a time span of about 5~µs.
The experimental sequence contains 12 operations, either read (\textbf{r}) or write (\textbf{w}).
We define three features which are necessary for use as a memory: a) that reading or writing to a rail should not affect its neighbors (interaction-free), b) rails which have not been written to should not return a retrieved pulse (empty state) and c) a read should leave the memory empty; a subsequent read pulse should yield no excitation (full retrieval).

Interaction-free operation is ensured by choosing an adequate rail separation and then confirmed by looking at the storage performance of a specific rail while there are operations performed on the neighboring rails.
The rails at 230~MHz (blue) and 190~MHz (green) can be used to show that operations are interaction free.
Between the write ($t=1.6~\textrm{µs}$) and read ($t=3.6~\textrm{µs}$) operation on the 190~MHz rail, four operations are performed on the neighboring rails and there is no visible impact on the read peak shape or height.
On the 230~MHz rail a pulse is written at $t=0$ and then retrieved during the last operation on the memory at $t=4.4~\textrm{µs}$.
Taking into account the 2.6~µs lifetime of this rail, the high remaining intensity of the retrieval peak clearly shows that there is no significant detrimental influence from multi-rail operation.

Reading an empty rail should not result in a significant amount of intensity.
We verify this by reading the 210~MHz and 170~MHz rail in a state that should not have excitation.
In the 210~MHz rail a pulse is written to the memory at $t=0.4~\textrm{µs}$ and then this rail is immediately read twice.
The second read operation at $t=0.8~\textrm{µs}$ yields negligible intensity compared to the first read operation at 0.6~µs.
Additionally the same rail is read again at 3.2~µs to observe the amount of noise, which is found to be comparable to the read at 1.2~µs.
The first operation on the 190~MHz rail at $t=2~\textrm{µs}$ is a read of a rail that has not been used before.
This allows us to determine how well the memory was initialized by the pumping that is performed prior to the experimental sequence.
Observing a larger intensity peak would point to insufficient polarization of the medium.
As the observed peak is similar to the other reads of an empty rail mentioned above, we conclude that pumping is sufficient and reading an empty rail, regardless of its history, does not lead to the detection of a significant peak.
In combination with the measurements on lifetime and rail interaction it follows that this setup allows random-access storage and retrieval of optical pulses for times comparable to the mean rail lifetime of 3.2~µs.

\section{Conclusion}
We have presented a multiplexed optical random-access memory, realised within a single vapor cell at a temperature of 60°C.
Using an EIT based storage scheme in a $\Lambda$-system on the cesium hyperfine transitions, we achieved a mean storage lifetime and internal efficiency of 3.2(2)~µs and 36\% respectively in multi-rail operation.
According to the chosen rail separation of 20~MHz, we performed random-access storage and retrieval in four parallel rails without observing reciprocal influence between the different rails.

The time between successive operations was chosen to be 400~ns for the sake of simplifying experiment control.
This time could be reduced considerably with the lower bound determined by the AODs switching time of 48~ns.
Increasing the storage lifetime and number of addressable rails is possible by increasing the beam diameters and using AODs that have a higher time-bandwidth product respectively.
This step will be important for applications in quantum communication and repeater networks.
Reaching beyond the threshold number of 1000 individually addressable modes is possible by using 2-axis AODs and multiplexing into a two dimensional grid of parallel storage modes.

\section*{Funding}
\noindent This work was funded by the Deutsche Forschungsgemeinschaft (DFG, German Research Foundation) – Project number 445183921.
E.R. acknowledges funding through the Helmholtz Einstein International Berlin Research School in Data Science (HEIBRiDS).

\section*{Disclosures}
\noindent The authors declare no conflicts of interest.

\section*{Data availability}
\noindent The data presented in this paper is available from the authors upon reasonable request.

\section*{References}
\bibliography{paper.bib}
\label{sec:refs}

\end{document}